\documentclass[twoside]{article}
\usepackage{graphicx,bm}
\usepackage[T2A]{fontenc}
\usepackage[cp1251]{inputenc}
\usepackage{cmpj2e}
\title[Two-loop RG functions]
{Two-loop RG functions
of the massive $\phi^4$ field theory in general dimensions}
\author{M. A. Shpot}
\address{Institute for Condensed Matter Physics, 79011 Lviv, Ukraine}
\newcommand{\ve}{\varepsilon}
\newcommand{\be}{\begin{equation}}
\newcommand{\ee}{\end{equation}}
\newcommand{\bea}{\begin{eqnarray}}
\newcommand{\eea}{\end{eqnarray}}
\begin{document}
\maketitle
\begin{abstract}
Two-loop Feynman integrals of the massive $\phi^4_d$ field theory
are explicitly obtained for generic space dimensions $d$.
Corresponding renormalization-group functions are expressed in a compact form
in terms of Gauss hypergeometric functions.
A number of interesting and useful relations is given for these integrals
as well as for several special mathematical functions and constants.
\pacs 11.10.Kk, 12.38.Bx, 02.30.Gp, 05.70.Jk
\end{abstract}

\section{Introduction}

The dependence of critical exponents of statistical mechanical systems on the
physical space dimension is one of the most fundamental features of
critical phenomena \cite{Fis74}. Continuous changing of the space dimension $d$
has been the basic principle for constructing the famous Wilson-Fisher epsilon
expansion \cite{WF72} with $\ve=4-d$.
However, the Renormalization Group (RG) \cite{WK74}, which
provides the fundamental theoretical basis for such calculations, is
nonperturbative in its nature. It does not require that the critical exponents
or any other universal quantities must be expanded in some space deviation,
coupling constant, inverse number of order-parameter components, etc.

While investigating the critical phenomena, the RG allows, in principle, to work
in any setting of interest provided that we are smart enough to find out
the appropriate analytical or numerical tools. The $\ve$-expansion has shown
itself as a very useful tool in studying the qualitative features of systems in
critical state even in its lowest-order approximations (see, e.g., \cite{DG76}).
Pushed to higher orders,
it allowed to produce very accurate numerical extrapolations
for critical exponents of three-dimensional $N$-vector models \cite{KSF}.
An alternative calculational scheme used with comparable success is the
so-called $g$-expansion within the massive field theory in fixed dimension
put forward by Parisi \cite{Par80}.
This approach has been mainly applied directly in three
dimensions to systems of different complexity
\cite{LZ77,LZ80,BSh92,DS98,PV00,PRV01}
(for a recent review see \cite{PV02}).
In \cite{LZ77} it was used for calculating the exponents of $\phi^4$ models in
two dimensions, and in \cite{HoSh92} --- applied to pure and disordered
Ising systems in general dimensions $2<d<4$.

A natural access to non-integer
dimensions provides also the well-known large-$N$ expansion
\cite{AH73,Ma73,Ma74JMP,MZ03}. In its general scope, it gives explicit expressions
for critical exponents as functions of $d$. These are valid in the whole range
between the lower and upper critical dimensionalities, and can be handled
analytically. The large-$N$ expansion is capable to yield information on
dimensional dependencies that are hardly accessible by other means, for example,
from the epsilon expansion.
Unfortunately, it is very hard to obtain such results in higher orders in $1/N$,
while short series expansions in $1/N$ usually fail to give very accurate
numerical estimates for relatively small values of $N$, say $N=3$.
The convergence of truncated $1/N$ expansions has been analyzed on the basis of
the field-theoretical approach at $d=3$ in \cite{AS95}.

The knowledge of dimensional dependencies of critical exponents, or eventually
also universal relations of critical amplitudes are of great theoretical and
practical interest. The information of this kind broadens out our fundamental
knowledge about these main characteristics of the critical behavior.
Moreover, the possibility to consider the results, which are not directly related
to dimensional expansions, in the vicinity of the upper or lower
critical dimensions provides us with useful checks of their correctness or with
some new relevant insights.

Let us give several examples. At the beginning of 1980ies Newman and Riedel
\cite{NR82} used the exact Wegner-Houghton \cite{WH73} RG equation and the scaling
field method in a study of the pure and dilute Ising models (IM)
in general dimensions from the interval $2.8<d<4$.

Pinn et. al. \cite{Pinn94} studied the RG in the hierarchical model in
$2<d<4$. Some of their results have been given as tables illustrating the
dimensional dependencies of critical indices calculated for some sets of
discrete non-integer values of $d$. A contact with the $\ve$-expansion has also
been established.

Recently, Ballhausen  et. al. \cite{BBW04} calculated dimensional dependencies
of IM critical exponents for $1<d<4$ by using the first-order derivative expansion
of the exact RG equation for the effective average action
(for recent reviews see \cite{BB01,BTW02}).
Their results for $\nu(d)$ and $\eta(d)$ are very similar to that obtained in
\cite{LZ87} and \cite{HoSh92}.

In a very recent paper \cite{DO08}, O'Dwyer and Osborn analyze the Polchinski
version \cite{Pol84} of the exact RG equations \cite{WH73,WK74,BB01}
and its possible truncations and
derivative expansions. This is done both for non-integer dimensions between two
and four, and in the epsilon expansion. Such combined approach allowed to better
understand the local potential approximation \cite{HH86}, the
derivative expansions used in treatments of the Polchinski equation, and moreover,
to suggest an alternative improved truncation of this equation.

Another story  happened in the study of the
critical behavior of ${O}(N)\times{O}(M)$ spin models
\cite{PRV01,Hol06c,Hol06r,Hol06a,Hol08}.
Apparently, certain controversies in the
discussion of these references could be avoided if there would be more
well-established information related to the dimensional dependencies of
RG functions, fixed points, and critical exponents.

Finally, let us note that there is a big continuous interest in the
high-energy physics (HEP) literature
in calculating explicit closed-form results for Feynman integrals
in general dimensions
\cite{BFT93,Tarasov96,DavDel98,DavKal01,FJT03,LapRem05,Tarasov06,%
ArgMast07,GKP07,Sh07,Tarasov08,KK09}.

We believe that the results displaying the explicit dependencies of
RG functions and physical observables on the space dimension $d$ are of
fundamental importance. The aim of the present paper is to do a little
step in this direction. We return to the article \cite{HoSh92} published
some time ago with Holovatch and calculate the closed-form expressions for the two-loop
RG functions of the massive field theory, treated in that work numerically,
in general dimensions $1\le d\le 4$.

\section{Two-loop RG functions and Feynman integrals}

Let us consider the usual massive \cite{Callan70,Sym70,Sym71,BLZ73,Par80}
$O(N)$-symmetric $\phi^4$ theory in general dimensions $d\le 4$.
To second order in renormalized
coupling constant $u$, the corresponding $\beta$ and $\gamma$ RG functions
can be written as \cite{BLZ73,Jug83,HoSh92}
\bea
&&\beta(u)=-(4-d)u\Big\{
1-u+\frac{8u^2}{(N+8)^2}\left[(5N+22)f(d)+(N+2)j(d)\right]\Big\}+O(u^4),
\nonumber\\&&\nonumber
\gamma_\phi(u)=-4(4-d)\frac{N+2}{(N+8)^2}\,j(d)\,u^2+O(u^3),
\nonumber\\&&\nonumber
\bar\gamma_{\phi^2}(u)=(4-d)\left[\frac{N+2}{N+8}\,u-12\frac{N+2}{(N+8)^2}\,f(d)\,u^2
\right]+O(u^3).
\eea
At the fixed point $u=u^*$, which is determined by the zero of the $\beta$ function
$\beta(u)$, the function $\gamma_\phi$ gives the value of the Fisher exponent $\eta$,
while $\bar\gamma_{\phi^2}(u^*)$ leads to the combination $2-\nu^{-1}-\eta$ where
$\nu$ is the correlation length critical exponent.

The $d$-dependent functions $f(d)\equiv i(d)-1/2$ and $j(d)$
are defined by the combinations of Feynman diagrams taken at zero external momenta,
\be\label{FJ}
i(d)=
\raisebox{-13pt}{\includegraphics[width=50pt]{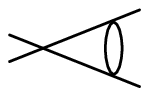}}\;/\;
\raisebox{-5pt}{\includegraphics[width=40pt]{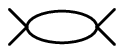}}\,^2\,
\quad\mbox{and}\quad
j(d)=\frac{\partial}{\partial q^2}\left.
\raisebox{-7pt}{\includegraphics[width=50pt]{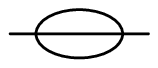}}\;\right |_{q^2=0}\;/
\raisebox{-5pt}{\includegraphics[width=40pt]{gamma4_1}}\,^2\,.
\ee
The massive two-loop integrals appearing in $i(d)$ and $j(d)$ are given by
\bea
&&I(m^2;d)=\int\int\frac{d^dk_1\,d^dk_2}{(2\pi)^{2d}}
\frac{1}{(k_1^2+m^2)^2(k_2^2+m^2)[(\bm k_1+\bm k_2)^2+m^2]}\,,
\\&&\label{JJ}
J(m^2,q^2;d)=\int\int\frac{d^dk_1\,d^dk_2}{(2\pi)^{2d}}
\frac{1}{(k_1^2+m^2)(k_2^2+m^2)[(\bm k_1+\bm k_2+\bm q)^2+m^2]}\,.
\eea
The one-loop integral which appears in (\ref{FJ}) and is traditionally used
\cite{BNGM76,NMB77,LZ77,Jug83} for normalization of the coupling constant is
given by
\be\label{DD}
D(m^2;d)=\int\frac{d^dk}{(2\pi)^d}\frac{1}{(k^2+m^2)^2}=m^{d-4}(4\pi)^{-d/2}\Gamma(2-d/2)
\ee
where $\Gamma(z)$ is a usual Euler's gamma function (see e.g. \cite{AS}).
This integral is absorbed, along with the symmetry factor $(N+8)/6$
into the normalization of the dimensionless renormalized coupling constant $u$.
With this normalization, the RG $\beta$ function starts with $-u+u^2+O(u^3)$
in the three-dimensional theory \cite{BNGM76,LZ77}. On the other hand, since
the integral $D(d)$ has a $1/\ve$ pole as $d$ approaches 4, in the epsilon expansion
the fixed point value $u^*$ will be given by $1+O(\ve)$
as a consequence of such normalization.

It is easy to see that the folowing relations hold true for the integrals
$I$ and $J$,
\be\label{DIN}
I(m^2;d)=-\frac{1}{3}\frac{\partial}{\partial m^2}J(m^2,0;d)
=-\frac{d-3}{3}\,(m^2)^{d-4}\,J(1,0;d),
\ee
where the second equality immediately follows from the scaling property
$J(m^2,q^2;d)=(m^2)^{d-3}J(1,q^2/m^2;d)$ of the function $J$.

In \cite{HoSh92}, using the Feynman parametrization (see, e.g., \cite{Ami84}),
$i(d)$ and $j(d)$ have been reduced to the double integrals
over Feynman parameters and written down as
\bea\label{FD}
&&i(d)=\frac{\Gamma(\ve)}{\Gamma^2(\ve/2)}\int_0^1\frac{x\,dx}{[x(1-x)]^{1-\ve/2}}
\int_0^1\frac{y^{\ve/2}\,dy}{[x(1-x)(1-y)+y]^\ve}\,,
\\\label{JD}&&
j(d)=-\frac{\Gamma(\ve)}{\Gamma^2(\ve/2)}\int_0^1\frac{dx}{[x(1-x)]^{-\ve/2}}
\int_0^1\frac{y^{\ve/2}(1-y)\,dy}{[x(1-x)(1-y)+y]^\ve}\,.
\eea
Here and further on, we parametrize the dimensional dependencies by the usual
deviation from the upper critical dimension $\ve=4-d$.
Generically, we do not assume that $\ve$ is infinitesimally small.

At the time of writing the paper \cite{HoSh92}, the following information was available.
At $d=3$, the integrals $i$ and $j$ could be easily evaluated analytically as
$i(3)=2/3$ and $j(3)=-2/27$.
At $d=2$, their numerical values are \cite{NMB77,Jug83}
$i(2)=0.78130241\ldots$ and $j(2)=-0.11463575\ldots$. As $d\to 4$, both integrals
$I(d)$ and $J(d)$ have poles in $\ve$, namely $I(4-\ve)\sim 1/\ve^2$ and
$J(4-\ve)\sim 1/\ve$.
In the functions $i(4-\ve)$ and $j(4-\ve)$ these  poles
are compensated by division through
$D^2(4-\ve)$, so that their $\ve$ expansions start with\footnote{
The similar formula given at p. 873 of \cite{HoSh92} is not correct.} \cite{BLZ73}
\be\label{EE}
i(d)=\frac{1}{2}+\frac{\ve}{4}+O(\ve^2)\quad\mbox{and}\quad
j(d)=-\frac{\ve}{8}-\frac{\ve^2}{8}\Big(\frac{3}{4}+I\Big)+O(\ve^3)
\ee
where
\be\label{BZLN}
I=\int_0^1 dx\left\{\frac{1}{1-x(1-x)}+\frac{\ln[x(1-x)]}{[1-x(1-x)]^2}\right\}.
\ee
This integral has not been calculated in \cite{BLZ73} since it is
"renormalization-dependent" and disappears from the results for the critical exponents
\cite{BLZ73}. Numerically, $I=-1.5626048\ldots$.
For non-integer $d$, the functions $f(d)$ and $j(d)$ have been tabulated
in \cite{HoSh92} by using numerical computations in (\ref{FD}) and (\ref{JD}).
The corresponding graphs have been plotted point by point for $2\le d\le 4$.
The aim of the following section is to obtain
analytically the explicit expressions for these functions in general dimensions
$d$.

\section{Explicit calculations in general dimensions}

Let us return to the integrals $i(d)$ and $j(d)$ given by (\ref{FD}) and (\ref{JD})
and consider the inner $y$ integral from (\ref{FD}).
Denoting for a while $z\equiv x(1-x)$ we write it as
\be\label{IF}
I_f(z)=z^{-\ve}\int_0^1 dy\, y^{\ve/2}[1-y(z-1)/z]^{-\ve}\,.
\ee
We recognize that here we have to deal with an integral representation of the
incomplete beta function $B_t(a,b)$ (see, e.g., \cite{AS}, entry 6.6.1).
But it is more convenient to express this function in terms of the Gauss hypergeometric
function using the relation 6.6.8 of \cite{AS},
\be\label{Bet}
B_t(a,b)=a^{-1}x^a\,_2F_1(a,1-b;a+1;t)\,.
\ee
Thus we obtain
\be
I_f(z)=\frac{2}{2+\ve}\;z^{-\ve}
\,_2F_1\left[1+\ve/2,\ve;2+\ve/2;(z-1)/z)\right]\,.
\ee
However, the argument of the resulting hypergeometric function is not good
for further integration over $x$ in the limits from $0$ to $1$
since the factor $1/z$ is singular at $x=0$ and $x=1$. The situation is substantially
improved by using the three-term linear transformation formula
15.38 from \cite{AS} which reads
\bea\nonumber
_2F_1(a,b;c;x)&=&\frac{\Gamma(c)\Gamma(b-a)}{\Gamma(b)\Gamma(c-a)}\;
(1-x)^{-a}\,_2F_1[a,c-b;a-b+1;(1-x)^{-1}]\\
&+&\mbox{a similar expression with $a$ and $b$ interchanged}.
\eea
After some algebraic transformations we obtain for the inner $y$ integral from
(\ref{FD})
\be\label{DW}
I_f(z)=\frac{2}{2-\ve}\left[-\frac{\ve}{2}\,\frac{\Gamma^2(\ve/2)}{\Gamma(\ve)}
\;z^{1-\ve/2}(1-z)^{-1-\ve/2}+\,_2F_1(\ve,1;\ve/2;z)\right]\,.
\ee
An essential simplification occurred in the first term due to an appearance
of a hypergeometric function with equal nominator and denominator parameters
(see, e.g., \cite{AS}, entry 15.1.8):
\be\label{HH}
_2F_1(a,b;b;z)=\sum_{k\ge 0}\frac{(a)_k}{k!}z^k=(1-z)^{-a}
\ee
where $(a)_k=\Gamma(a+k)/\Gamma(a)=a(a+1)\ldots(a+k-1)$
is the Pochhammer symbol.

Let us substitute the first term of (\ref{DW}) into the outer $x$ integral
remaining in (\ref{FD}). Recalling the short-hand notation $z=x(1-x)$ we see
that the factor $z^{1-\ve/2}$ is exactly cancelled by the analogous term in
the denominator and, apart from the numerical constants, we remain with
\be\label{KK}
\int_0^1 \frac{x\,dx}{(1-x+x^2)^{1+\ve/2}}=
\int_0^{1/2}\frac{dt}{(\frac{3}{4}+t^2)^{1+\ve/2}}=
\frac{2^{1+\ve}}{3^{1+\ve/2}}\;
_2F_1\Big(1+\frac{\ve}{2},\frac{1}{2};\frac{3}{2};-\frac{1}{3}\Big)\,.
\ee
The last equality can be obtained by simply expanding the denominator
of the $t$ integral via (\ref{HH})
and integrating the resulting series expansion term by term.

The second part of the square brackets in (\ref{DW}) can be handled in a similar
fashion. A straightforward calculation employing the Gauss series representation
of $_2F_1$ yields
\be\label{PP}
\int_0^1\frac{x\,dx}{[x(1-x)]^{1-\ve/2}}
\;_2F_1\Big[\ve,1;\frac{\ve}{2};x(1-x)\Big]
=\frac{\Gamma^2(\ve/2)}{2\Gamma(\ve)}\;
_2F_1\Big(\ve,1;\frac{1}{2}+\frac{\ve}{2};\frac{1}{4}\Big)\,.
\ee
Now we notice that the hypergeometric function with the argument $-1/3$ can be
expressed in terms of a similar function with the argument $1/4$ using the
the linear transformation formula (see, e.g., \cite{AS}, entry 15.3.4)
\begin{equation}\label{Ltran}
_2F_1(a,b;c;z)=(1-z)^{-a}\,_2F_1\Big(a,c-b;c;\frac{z}{z-1}\Big)\,.
\end{equation}
Thus we get from (\ref{FD}) and (\ref{KK}) along with (\ref{Ltran})
and (\ref{PP}) our first explicit result
\be\label{SS}
i(d)=\frac{1}{2-\ve}\left[
\,_2F_1\Big(\ve,1;\frac{1}{2}+\frac{\ve}{2};\frac{1}{4}\Big)
-\frac{\ve}{2}\;
_2F_1\Big(1+\frac{\ve}{2},1;\frac{3}{2};\frac{1}{4}\Big)\right]\,.
\ee
As $\ve\to 0$, the first hypergeometric function in square brackets
reduces to $1$, the second term vanishes due to the explicit factor $\ve$,
and $i(d)=1/2+O(\ve)$, as it should
(see (\ref{EE})). Despite of the overall factor $1/(2-\ve)$,
there is no singularity at $d=2$ since the combination
in square brackets vanishes at $\ve=2$. In Appendix we derive the explicit
result for $i(2)$ in terms of the generalized hypergeometric function $_3F_2$.

Successive application of Gauss' relations for contiguous functions $_2F_1$
given by entries 7.3.1.16 and 7.3.1.18 of \cite{PBM3} to the second hypergeometric
function from (\ref{SS}) leads to an elegant representation
\be\label{SF}
i(d)=\frac{1}{2-\ve}\left[2+
\,_2F_1\Big(\ve,1;\frac{1}{2}+\frac{\ve}{2};\frac{1}{4}\Big)
-2\;_2F_1\Big(\frac{\ve}{2},1;\frac{1}{2};\frac{1}{4}\Big)\right]\,.
\ee
Here an interesting symmetry of the nominator and denominator
parameters in both functions $_2F_1$ is observed, while they differ one from another
by $\ve/2$.

Let us consider the second function $j(d)$.
The calculation of the double integral (\ref{JD}) is somewhat more involved but the
experience of the above calculation allows us to be brief.

While considering the inner integration $I_f(z)$ in (\ref{IF}) we have seen
that it is useful to get its result in terms of a Gauss hypergeometric function
of the argument $z\equiv x(1-x)$, and not $(z-1)/z$ which is suggested directly
by the integrals over $y$ in (\ref{FD})--(\ref{JD}) and (\ref{IF}).
This can be achieved directly on the level of the integral representation.
Indeed, changing the initial integration variable via
$y=1/(t+1)$ we obtain for the inner integral $I_j(z)$ from (\ref{JD})
\bea\label{DW1}
I_j(z)&=&\int_0^\infty\frac{t\,dt}{(t+1)^{3-\ve/2}}\,(1+zt)^{-\ve}=
\frac{4}{(4-\ve)(2-\ve)}\,_2F_1(\ve,2;\ve/2;z)
\\\nonumber
&-&\frac{\ve}{(4-\ve)(2-\ve)}
\,\frac{\Gamma^2(\ve/2)}{\Gamma(\ve)}\;z^{1-\ve/2}(1-z)^{-2-\ve/2}
\left[4-\ve-2(1-\ve)z\right]\,.
\eea

The two terms of $I_j(z)$ are similar to that generated by the inner
$y$ integration in $i(d)$ (see (\ref{DW})). They can be handled in the same
manner inside of the external integral over $x$ in (\ref{JD}).
According to this decomposition of $I_j(z)$, we obtain two different contributions
to the function $j(d)$. We can write
\be
j(d)=-\frac{\ve}{(4-\ve)(2-\ve)}\left[j_1(d)+j_2(d)\right]
\ee
where
\bea\nonumber
&&j_1(d)=\frac{1-\ve}{3(1{+}\ve)}\left[
_2F_1\Big(\ve,1;\frac{3{+}\ve}{2};\frac{1}{4}\Big)\right.
+\left.\frac{4(4+\ve)}{3(1-\ve)}-\frac{4+\ve}{3(3{+}\ve)}\,
_2F_1\Big(1+\ve,1;\frac{5{+}\ve}{2};\frac{1}{4}\Big)\right],
\\\nonumber&&
j_2(d)=\frac{2^{1+\ve}}{3^{\ve/2}}\left[(1-\ve)
_2F_1\Big(\frac{\ve}{2},\frac{1}{2};\frac{3}{2};-\frac{1}{3}\Big)
+2\ve\,_2F_1\Big(1+\frac{\ve}{2},\frac{1}{2};\frac{3}{2};-\frac{1}{3}\Big)\right.
\\\nonumber&&\hspace{25mm}
-\frac{8}{9}(2+\ve)\left.
\,_2F_1\Big(2+\frac{\ve}{2},\frac{1}{2};\frac{3}{2};-\frac{1}{3}\Big)\right].
\eea

Similarly as before, using the linear transformation
(\ref{Ltran}), we can reduce all above hypergeometric functions to that
with the argument $1/4$. Further, the repeated use of the contiguous functions'
relations leaves us only with the same two $_2F_1$ functions
which appeared in $i(d)$. Moreover, they build up just the combination defining
that function. Finally we arrive at a simple expression of $j(d)$ in terms of $i(d)$:
\be\label{JFR}
j(d)=\frac{1}{3d}\Big[4-(4+d)i(d)\Big].
\ee
We recall that two equivalent explicit expressions for the function $i(d)$ are
given in (\ref{SS}) and (\ref{SF}).

\section{Some results in two dimensions}\label{2D}

While the integrals $i(d)$ and $j(d)$ are smooth functions of $d$ in the range
of our interest, their calculation at $d=2$ requires some care. In order to
calculate $i(2)$ from (\ref{SS}) or  (\ref{SF}), we have to introduce a small deviation
$\alpha$ from $d=2$ and consider the limits $\alpha\to 0$ of these expressions.
Details of the calculation can be found in appendix \ref{App1}.

We obtain the explicit expression for $i(d)$ at $d=2$ in terms of the generalized
hypergeometric function $_3F_2$:
\be\label{Fex2}
i(2)=\frac{4}{3\sqrt 3}\,
_3F_2\Big(\frac{1}{2},\frac{1}{2},\frac{1}{2};
\frac{3}{2},\frac{3}{2};\frac{1}{4}\Big)\,.
\ee
Using the relation (\ref{JFR}) it is easy to obtain the value of $j(2)$:
\be
j(2)=-\frac{4}{3\sqrt 3}\,
_3F_2\Big(\frac{1}{2},\frac{1}{2},\frac{1}{2};
\frac{3}{2},\frac{3}{2};\frac{1}{4}\Big)+\frac{2}{3}\,.
\ee

Searching for some other representations of $i(2)$ we have found the relation
(details are given in appendix \ref{App1})
\be\label{FCR}
_3F_2\Big(\frac{1}{2},\frac{1}{2},\frac{1}{2};\frac{3}{2},\frac{3}{2};\frac{1}{4}\Big)
=\mbox{Cl}_2(\pi/3)\,.
\ee
Here $\mbox{Cl}_2(\pi/3)=1.0149417\ldots$ is the maximum value of the Clausen's
function (for more information on this and related functions see \cite{Lewin})
\be\label{CL2d}
\mbox{Cl}_2(\theta)=-\int_0^\theta d\theta\ln\Big|2\sin\frac{\theta}{2}\Big|
=\sum_{n\ge1}\frac{\sin n\theta}{n^2}\,.
\ee
Apparently, the identity (\ref{FCR}) does not appear explicitly in the mathematical
literature. It supplements the very similar formula
\be\label{FCP1}
_3F_2\Big(\frac{1}{2},\frac{1}{2},\frac{1}{2};\frac{3}{2},\frac{3}{2};-\frac{1}{4}\Big)
=\frac{\pi^2}{10}
\ee
from mathematical tables \cite{PBM3}, entry 7.4.6.1.

\section{What can we learn from the HEP literature?}
As well as in the condensed-matter physics,
the calculation of multi-loop Feynman integrals is an important issue in the
high-energy physics literature. Although the Feynman diagrams in four
space-time dimensions are of main concern in the HEP theory, very often they
are calculated by using the dimensional regularization in arbitrary space-time
dimension $d$. It is recognized (see e.g. \cite{BFT93}) that
working in general $d$ dimensions is often much easier than directly at $d=4$.

Actually, the condensed-matter and
high-energy physicists are often doing much the same work
by pursuing physically different goals.
This is, perhaps, most pronounced just in calculations of Feynman integrals.
Unfortunately, it happens not very often that authors of these two communities
pay enough attention to achievements gained in another one. The aim of the present
section is to quote some HEP results related to calculations of the preceding sections.
In doing this we show several useful implications of establishing such contacts.
By reviewing briefly some HEP references we would like to draw attention
of the cond-mat readers to them, as we feel that these are not well known and
appreciated in the cond-mat literature.

Some of the results obtained in \cite{BFT93} are directly related to ours.
In particular, the authors of this reference calculate the two-point function
$J_3(m^2,q^2)\sim\raisebox{-8pt}{\includegraphics[width=48pt]{gamma2_3}}$.
By using the Mellin-Barnes contour integral
representation for massive propagators \cite{BD91} they derive for $J_3(m^2,q^2)$
an infinite series over linear combinations of $_3F_2$ functions\footnote{
Only quite recently an explicit expression for $J_3(m^2,q^2)$ has been found
\cite{Tarasov06} in terms of Gauss and Appell hypergeometric functions.}.

The first two coefficients of the small-$q^2$ expansion of $J_3(1,q^2)$ are simply
proportional to our functions $i(d)$ and $j(d)$. This can be seen by noticing
(see (\ref{JJ})-(\ref{DD})) that
$J(m^2,q^2)/D^2(m^2)=(\ve^2/4)J_3(m^2,q^2)$ and taking into account the relation
(\ref{DIN}). Thus we get, first,
\be\label{BFT1}
i(d)=\frac{1}{2-\ve}\,\frac{2}{3}\,C(0)=\frac{1}{2-\ve}\,\frac{2}{3}\left[
3^{\frac{1}{2}-\frac{\ve}{2}}\,\frac{\pi\Gamma(\ve)}{\Gamma^2(\ve/2)}
+\frac{3}{2}(1-\ve)\,_2F_1\Big(\frac{\ve}{2},1;\frac{3}{2};\frac{1}{4}\Big)\right]
\ee
where we quote the equation (33) from \cite{BFT93} with the usual replacement
$\ve\to\ve/2$. Analogous expressions can be found also in \cite{DavKal01}, eq. (4.38),
and \cite{PV00np}, eq. (A.11).

The last expression for $i(d)$ is equivalent to that found in (\ref{SS}) and (\ref{SF}).
A direct comparison between (\ref{BFT1}) and (\ref{SS})-(\ref{SF}) can be done by
applying the contiguous relations for hypergeometric functions with $\ve/2$
in nominator parameters. This leads to an interesting and non-trivial
relation for involved Gauss' functions differing in $O(\ve)$ as $\ve\to 0$,
\be\label{XAX}
_2F_1\Big(\ve,1;\frac{1}{2}+\frac{\ve}{2};\frac{1}{4}\Big)+\,
_2F_1\Big(\frac{\ve}{2},1;\frac{1}{2};\frac{1}{4}\Big)-2
=3^{-\frac{1}{2}-\frac{\ve}{2}}\,\frac{2\pi\Gamma(\ve)}{\Gamma^2(\ve/2)}\,.
\ee

The second check is given by the relation
\be\label{BSS}
j(d)=\frac{4}{d(d{-}2)(d{-}4)}\,C(1)
\ee
where $C(1)$ is the small-$q^2$ expansion coefficient
appearing in eq. (31) of \cite{BFT93}.
This coefficient is expressible in terms of $C(0)$ by the recurrence relation (37)
of \cite{BFT93},
\be\label{BSF}
C(1)=\frac{d-3}{3}\Big[ d-2 - \frac{d+4}{6}\,C(0) \Big].
\ee
Now, taking into account the first equality of (\ref{BFT1}), we reproduce
from (\ref{BSS})-(\ref{BSF}) the previously obtained expression
(\ref{JFR}) for $j(d)$ in terms of $i(d)$.
Thus, the formula (\ref{JFR}) is a consequence of recurrent relations that
hold true for the small-$q^2$ expansion coefficients of the Feynman integral
$J(m^2,q^2;d)$. In turn, these recurrences follow from the differential equation%
\footnote{For a recent review on applications of differential equations
in calculations of Feymnan integrals see \cite{ArgMast07}.}
satisfied by the function $J(m^2,q^2;d)$ \cite{BFT93}.

A full $\ve$ expansion of the functions $i(d)$ and $j(d)$ can be obtained with the help
of eq. (4.16) from \cite{DavKal01}. Using it we can write
\be\label{EQ33}
i(d)=\frac{1}{2-\ve}\left\{1+
3^{-\frac{1}{2}-\frac{\ve}{2}}\,\frac{2\pi\Gamma(\ve)}{\Gamma^2(\ve/2)}
-\frac{\ve}{2}\,3^{\frac{1}{2}-\frac{\ve}{2}}
\sum_{j\ge 0}\frac{\ve^j}{j!}\Big[\mbox{Ls}_{j+1}\Big(\frac{2\pi}{3}\Big)-
\mbox{Ls}_{j+1}(\pi)\Big]\right\}
\ee
where
\be\label{LJD}
\mbox{Ls}_j(\theta)=-\int_0^\theta d\theta\ln^{j-1}\Big|2\sin\frac{\theta}{2}\Big|
\ee
are the log-sine functions (see \cite{DavKal01,Lewin}). Per definition,
\be\label{PRT}
\mbox{Ls}_1(\theta)=-\theta \quad\mbox{and}\quad \mbox{Ls}_2(\theta)=\mbox{Cl}_2(\theta)
\ee
where $\mbox{Cl}_2(\theta)$ is the Clausen's function. This special function appeared
readily in Sec. \ref{2D}, in the calculation of the integrals $i(d)$ and $j(d)$
in two dimensions. With the first two terms from the sum, we get the $\ve$ expansions
\be\label{LDE}
i(d)=\frac{1}{2}+\frac{\ve}{4}+\frac{\ve^2}{8}-
\frac{\ve^2}{2\sqrt 3}\,\mbox{Cl}_2\Big(\frac{\pi}{3}\Big)+O(\ve^3)
\ee
and
\be\label{JDE}
j(d)=-\frac{\ve}{8}-\frac{3\ve^2}{32}+
\frac{\ve^2}{3\sqrt 3}\,\mbox{Cl}_2\Big(\frac{\pi}{3}\Big)+O(\ve^3)\,.
\ee
In deriving these we took into account that
$\mbox{Cl}_2(2\pi/3){=}2/3{\cdot}\mbox{Cl}_2(\pi/3)$, $\mbox{Cl}_2(\pi){=}0$, and
used the relation (\ref{JFR}) between $i(d)$ and $j(d)$. The non-trivial $O(\ve^2)$
term appearing in curly brackets of (\ref{EQ33}) agrees with that of
equation (33) in \cite{BFT93}.

We see that a very special transcendental constant $\mbox{Cl}_2(\pi/3)$,
which is the maximum value of the Clausen's integral $\mbox{Cl}_2(\theta)$,
appears both in the epsilon expansion of $i(d)$ near $d=4$ and in the
explicit expression of $i(d)$ at $d=2$. This is quite natural because
the Feynman integrals of the type considered here obey certain general relations
that connect their values in different space dimensions
\cite{DavTausk96,Tarasov96,LapRem05,Tarasov06}.

A simple formula relating the values of Feynman integrals associated with the
zero-momentum "sunrise" diagram $\raisebox{-7pt}{\includegraphics[width=40pt]{gamma2_3}}$
with arbitrary masses on the lines at $d=4-\ve$ and $d=2-\ve$ has been found in
\cite{DavTausk96}. In the particular equal-mass case with $m^2=1$, this formula reads,
in our notation,
\be
J(1,0;4-\ve)=\frac{3\pi^2}{(1{-}\ve)(2{-}\ve)}\Big[J(1,0;2-\ve)-\pi^{2-\ve}
\Gamma^2\Big(\frac{\ve}{2}\Big)\Big]\,.
\ee
Note that the pole terms of the Laurent expansion of $J(4-\ve)$ are contained
only in the last, trivial term, and the value $J(2)$ directly enters the finite part
of $J(4-\ve)$ (cf. \cite{LapRem05}, Sec. 3-4).
This feature directly maps onto the connection between the
functions $i(4-\ve)$ and $i(2-\ve)$, which can be written as
\be
i(4-\ve)=\frac{1}{2-\ve}\Big[1-\frac{3\ve^2}{4(1+\ve)}\,i(2-\ve)\Big]\,.
\ee
It is straightforward to check that the last relation is indeed satisfied by the function
$i(d)$ given, for instance, by (\ref{BFT1}).

Let us return to the epsilon expansions (\ref{LDE})--(\ref{JDE}).
They can be compared with that quoted in (\ref{EE})--(\ref{BZLN}).
In (\ref{LDE}) we see the $O(\ve^2)$ term of $i(d)$, which is missing in (\ref{EE}).
By comparing the $O(\ve^2)$ terms of $j(d)$ we can make the identification
\be\label{BZN}
I=\int_0^1 \frac{dx}{1-x(1-x)}\left\{1+\frac{\ln[x(1-x)]}{[1-x(1-x)]}\right\}
=-\frac{8}{3\sqrt 3}\,\mbox{Cl}_2\Big(\frac{\pi}{3}\Big)\,.
\ee
An attempt to calculate this integral with the help of Mathematica \cite{math}
yields an expression in terms of $\psi^\prime(x)$ where $\psi(x)$ is the psi function,
the logarithmic derivative of the Gamma function (see e.g. \cite{PBM1}). Thus we get
the chain of equalities
\be\label{FCP}
_3F_2\Big(\frac{1}{2},\frac{1}{2},\frac{1}{2};\frac{3}{2},\frac{3}{2};\frac{1}{4}\Big)
=\mbox{Cl}_2\Big(\frac{\pi}{3}\Big)=\frac{1}{2\sqrt 3}\;\psi^\prime\Big(\frac{1}{3}\Big)-
\frac{\pi^2}{3\sqrt 3}\,.
\ee
The rightmost connection can be also read off from (A.13) of \cite{PV00}.

\section{Concluding remarks}
Let us return to the two-loop RG functions defined at the beginning of the Section 2.
We see that they can be expressed in terms of a single function of $d=4-\ve$, say
\be
i(d)=\frac{1}{2-\ve}\left[4-3\,_2F_1\Big(\frac{\ve}{2},1;\frac{1}{2};\frac{1}{4}\Big)+
3^{-\frac{1}{2}-\frac{\ve}{2}}\,\frac{2\pi\Gamma(\ve)}{\Gamma^2(\ve/2)}\right],
\ee
which contains only one non-trivial Gauss' hypergeometric function.
For example, the function $\gamma_\phi(u)$ is thus given by
\be
\gamma_\phi(u)=-4\,\frac{4-d}{3d}\frac{N+2}{(N+8)^2}\,\Big[4-(4+d)i(d)\Big]\,u^2+O(u^3).
\ee
Similar expressions can be constructed also for the remaining RG functions $\beta(u)$
and $\bar\gamma_{\phi^2}(u)$.

Everywhere the dependence on space dimension $d$ is given explicitly in a simple
parametric form. Obviously, the convenience of such expressions is much better compared
to representations in terms of multiple integrals as that in (\ref{FD})--(\ref{JD}).
In principle, this kind of formulas
could be used in considering some dimensional expansions not tied to $d=4$.

Analytical two-loop results are interesting in view of availability of numerical
tables of three-loop diagrams of the massive field theory in $d$ dimensions \cite{HK94}.
It could be thought of extending explicit calculations to that order.

Finally, they are also interesting in their own right, as any mathematical
results derived in closed form. Moreover, in the course of the present work
it was possible to find out some interesting mathematical relations given in
equations (\ref{FCR}), (\ref{FCP}) and (\ref{XAX}).

\appendix
\section{Explicit results in two dimensions}\label{App1}
In order to derive the values of the functions $i(d)$ and $j(d)$ at $d=2$
we have to take $\ve=2-\alpha$ and consider the corresponding limits with
$\alpha\to 0$. Thus we write $i(d)$ in (\ref{SS}) as
\be
i(d)=i_1(\alpha)+i_2(\alpha)
\ee
with
\be
i_1(\alpha)=\frac{1}{\alpha}\left[
\,_2F_1\Big(2-\alpha,1;\frac{3}{2}-\frac{\alpha}{2};\frac{1}{4}\Big)-\;
_2F_1\Big(2-\frac{\alpha}{2},1;\frac{3}{2};\frac{1}{4}\Big)\right]
\ee
and
\be
i_2(\alpha)=\frac{1}{2}
\,_2F_1\Big(2-\frac{\alpha}{2},1;\frac{3}{2};\frac{1}{4}\Big)\,.
\ee
There exists a simple $\alpha\to 0$ limit of $i_2(\alpha)$,
\be\label{Fe2}
i_2(0)=\frac{1}{3}+\frac{2\pi}{9\sqrt 3}\,,
\ee
while each of the hypergeometric functions in $i_1(\alpha)$ has to be expanded
to $O(\alpha)$ in order to give the finite $i_1(0)$. To this end, we found it
useful to transform these functions via the linear transformation formula
(\cite{AS}, entry 15.3.3)
\begin{equation}\label{L4t}
_2F_1(a,b;c;z)=(1-z)^{c-a-b}\,_2F_1\Big(c-a,c-b;c;z\Big)\,.
\end{equation}
Thus we get
\be\nonumber
i_1(\alpha)=\Big(\frac{3}{4}\Big)^{-3/2+\alpha/2}\frac{1}{\alpha}\left[
\,_2F_1\Big(-\frac{1}{2}+\frac{\alpha}{2},\frac{1}{2}-\frac{\alpha}{2};
\frac{3}{2}-\frac{\alpha}{2};\frac{1}{4}\Big)-\,
_2F_1\Big(-\frac{1}{2}+\frac{\alpha}{2},\frac{1}{2};\frac{3}{2};\frac{1}{4}\Big)\right].
\ee
Here the both Gauss functions are of the type (\ref{Bet}), and hence can be represented
in terms of simple integrals appearing in (\ref{IF}). We have
\be\label{F1al}
i_1(\alpha)=\Big(\frac{3}{4}\Big)^{-3/2+\alpha/2}\frac{1}{2\alpha}\left[
(1-\alpha)I_1-I_2\right]=\Big(\frac{3}{4}\Big)^{-3/2+\alpha/2}\frac{1}{2\alpha}\left[
(I_1-I_2)-\alpha I_1\right]
\ee
where
\be
I_1=\int_0^1 dt\,t^{-\frac{1}{2}-\frac{\alpha}{2}}(1-t/4)^{\frac{1}{2}-\frac{\alpha}{2}}
\quad\mbox{and}\quad
I_2=\int_0^1 dt\,t^{-\frac{1}{2}}(1-t/4)^{\frac{1}{2}-\frac{\alpha}{2}}.
\ee
At $\alpha=0$, the integrals $I_1$ and $I_2$ both are equal to
\be\label{Ia0}
I_0=\frac{\pi}{3}+\frac{\sqrt 3}{2}\,,
\ee
while the difference between them $I_1-I_2$ is of order $\alpha$:
\be
I_1-I_2=
\int_0^1 \frac{dt}{\sqrt t}(1-t/4)^{\frac{1}{2}-\frac{\alpha}{2}}
(t^{-\frac{\alpha}{2}}-1)=
-\frac{\alpha}{2}\int_0^1 \frac{dt}{\sqrt t}(1-t/4)^\frac{1}{2}\ln t +O(\alpha^2).
\ee
Using Mathematica \cite{math} we get
\be\label{Ri12}
I_1-I_2=\alpha\left[\frac{\pi}{6}+\frac{\sqrt 3}{4}+
\,_3F_2\Big(\frac{1}{2},\frac{1}{2},\frac{1}{2};
\frac{3}{2},\frac{3}{2};\frac{1}{4}\Big)\right]+O(\alpha^2).
\ee
Combining (\ref{F1al}), (\ref{Ri12}), and (\ref{Ia0}) we obtain
\be\label{Plast}
i_1(0)=-\frac{2\pi}{9\sqrt 3}-\frac{1}{3}+\frac{4}{3\sqrt 3}\,
_3F_2\Big(\frac{1}{2},\frac{1}{2},\frac{1}{2};
\frac{3}{2},\frac{3}{2};\frac{1}{4}\Big)\,,
\ee
and finally, together with $i_2(0)$ from (\ref{Fe2}), we arrive at $i(2)$ given
in equation (\ref{Fex2}) of the main text.

We did not find in mathematical tables any summation formula for
the generalized hypergeometric function $_3F_2$
appearing in the last two equations. So we looked for the series
representations of this function
\be\label{Dser}
\sum_{k\ge 0}\frac{(1/2)_k}{k!(2k+1)^2}\frac{1}{4^k}=
4\sum_{k\ge 0}\frac{(2k)!}{(k!)^2(2k+1)^2}\Big(\frac{1}{4}\Big)^{2k+1}=
\sum_{k\ge 0}\frac{(2k-1)!!}{(2k)!!(2k+1)^2}\frac{1}{4^k}\,.
\ee
The middle version of the series appears, indeed, as a partial case of
the formula 5.2.13.7 in \cite{PBM1} expressed in terms of the function $\arcsin 2x$
and its derivative. But, unfortunately, a simple numerical check at $x=1/4$
revealed the incorrectness of this entry.

In order to find the right answer, we considered the series expansion
\be
g(x)\equiv\sum_{k\ge 0}\frac{(1/2)_k}{k!(2k+1)^2}\frac{x^{2k+1}}{4^k}\,,
\ee
such that $g(1)=\,_3F_2(1/2,1/2,1/2;3/2,3/2;1/4)$ and $g(0)=0$.
Its derivative with respect to $x$, multiplied by $x$, is
\be
x\,g'(x)=\sum_{k\ge 0}\frac{(1/2)_k}{k!(2k+1)}\frac{x^{2k+1}}{4^k}\,.
\ee
Differentiating once again we get
\be
\left[x\,g'(x)\right]'=\sum_{k\ge 0}\frac{(1/2)_k}{k!}\frac{x^{2k}}{4^k}
=\frac{1}{\sqrt{1-x^2/4}}\,.
\ee
Integrating back leads us to
\be
x\,g'(x)=2\arcsin\frac{x}{2}
\ee
and
\be
g(x)=\int dx\frac{2}{x}\arcsin\frac{x}{2}=
\mbox{Cl}_2\big(2\arcsin\frac{x}{2}\big)+\arcsin\frac{x}{2}\,\ln x
\ee
where $\mbox{Cl}_2(z)$ is the Clausen function \cite{Lewin}.
In deriving the last equality we used the entry 1.7.4.13 from \cite{PBM1}
and the fact that the integration constants in the both last indefinite integrations
vanish. Here we observe also that the neighboring entry 1.7.4.12 of \cite{PBM1}
involves the series expansion of the same type as the rightmost one in (\ref{Dser}),
giving, along with 1.7.4.11, the corrected version of the false formula 5.2.13.7 in
the came reference.

Now, recalling that at $x=1$, the function $g(x)$ reproduces the hypergeometric series
$_3F_2$ from (\ref{Plast}), we come to the relation (\ref{FCR}) given in the main text.


\begin{thebibliography}{10}
\def\selectlanguageifdefined#1{
\expandafter\ifx\csname date#1\endcsname\relax
\else\language\csname l@#1\endcsname\fi}
\ifx\undefined\url\def\url#1{#1}\else\fi

\bibitem{Fis74}
\selectlanguageifdefined{english}
{Fisher~Michael~E.} The renormalization group in the theory of critical
  behavior~// {Rev. Mod. Phys.} ---
\newblock 1974. ---
\newblock Vol.~46, №~4. ---
\newblock Pp.~597--616.

\bibitem{WF72}
\selectlanguageifdefined{english}
{Wilson~K.~G., Fisher~M.~E.} Critical exponents in 3.99 dimensions~// {Phys.
  Rev. Lett.} ---
\newblock 1972. ---
\newblock Vol.~28, №~4. ---
\newblock Pp.~240--243.

\bibitem{WK74}
\selectlanguageifdefined{english}
{Wilson~Kenneth~G., Kogut~J.} The renormalization group and the $\epsilon$
  expansion~// {Phys. Reports}. ---
\newblock 1974. ---
\newblock Vol. 12C, №~2. ---
\newblock Pp.~75--200.

\bibitem{DG76}
\selectlanguageifdefined{english}
{Domb~C., Green~M.~S.}, eds. Phase Transitions and Critical Phenomena. ---
\newblock London: Academic Press, 1976. ---
\newblock Vol.~6.

\bibitem{KSF}
\selectlanguageifdefined{english}
{Kleinert~H., Schulte-Frohlinde~V.} Critical Properties of $\phi^4$-Theories.
  ---
\newblock Singapore: World Scientific, 2000.

\bibitem{Par80}
\selectlanguageifdefined{english}
{Parisi~Giorgio}. Field-theoretic approach to second-order phase transitions in
  two- and three-dimensional systems~// {J. Stat. Phys.} ---
\newblock 1980. ---
\newblock Vol.~23, №~1. ---
\newblock Pp.~49--81.

\bibitem{LZ77}
\selectlanguageifdefined{english}
{Le~Guillou~J.~C., Zinn-Justin~J.} Critical exponents for the $n$-vector model
  in three dimensions from field theory~// {Phys. Rev. Lett.} ---
\newblock 1977. ---
\newblock Vol.~39, №~2. ---
\newblock Pp.~95--98.

\bibitem{LZ80}
\selectlanguageifdefined{english}
{Le~Guillou~J.~C., Zinn-Justin~J.} Critical exponents from field theory~//
  {Phys. Rev. B}. ---
\newblock 1980. ---
\newblock Vol.~21, №~9. ---
\newblock Pp.~3976--3998.

\bibitem{BSh92}
\selectlanguageifdefined{english}
{Bervillier~C., Shpot~M.} Universal amplitude combinations of the
  three-dimensional random {I}sing system~// {Phys. Rev. B}. ---
\newblock 1992. ---
\newblock Vol.~46, №~2. ---
\newblock Pp.~955--968.

\bibitem{DS98}
\selectlanguageifdefined{english}
{Diehl~H.~W., Shpot~M.} Massive field-theory approach to surface critical
  behavior in three-dimensional systems~// {Nucl. Phys. B}. ---
\newblock 1998. ---
\newblock Vol. 528, №~3. ---
\newblock Pp.~595--647.

\bibitem{PV00}
\selectlanguageifdefined{english}
{Pelissetto~Andrea, Vicari~Ettore}. Randomly dilute spin models: A six-loop
  field-theoretic study~// {Phys.\ Rev.\ B}. ---
\newblock 2000. ---
\newblock Vol.~62, №~10. ---
\newblock Pp.~6393--6409.

\bibitem{PRV01}
\selectlanguageifdefined{english}
{Pelissetto~Andrea, Rossi~Paolo, Vicari~Ettore}. Large-$n$ critical behavior of
  ${O}(n)\times{O}(m)$ spin models~// {Nucl. Phys. B}. ---
\newblock 2001. ---
\newblock Vol. 607, №~3. ---
\newblock Pp.~605--634.

\bibitem{PV02}
\selectlanguageifdefined{english}
{Pelissetto~A., Vicari~E.} Critical phenomena and renormalization group
  theory~// {Phys. Rep.} ---
\newblock 2002. ---
\newblock Vol. 368. ---
\newblock Pp.~549--727.

\bibitem{HoSh92}
\selectlanguageifdefined{english}
{Holovatch~{Yu}., Shpot~M.} Critical exponents of random {I}sing-like systems
  in general dimensions~// {J. Stat. Phys.} ---
\newblock 1992. ---
\newblock Vol.~66, no. 3/4. ---
\newblock Pp.~867--883.

\bibitem{AH73}
\selectlanguageifdefined{english}
{Abe~Ryuzo, Hikami~Shinobu}. Critical exponents and scaling relations in $1/n$
  expansion~// {Progr. Theor. Phys.} ---
\newblock 1973. ---
\newblock Vol.~49, №~2. ---
\newblock Pp.~442--452.

\bibitem{Ma73}
\selectlanguageifdefined{english}
{Ma~{S-k}.} Critical exponents above ${T}_c$ to ${O}(1/n)$~// {Phys.\ Rev.\ A}.
  ---
\newblock 1973. ---
\newblock Vol.~7, №~6. ---
\newblock Pp.~2172--2187.

\bibitem{Ma74JMP}
\selectlanguageifdefined{english}
{Ma~{S-k}.} The renormalization group and the large $n$ limit~// {J. Math.
  Phys.} ---
\newblock 1974. ---
\newblock Vol.~15, №~11. ---
\newblock Pp.~1866--1891.

\bibitem{MZ03}
\selectlanguageifdefined{english}
{Moshe~Moshe, Zinn-Justin~Jean}. Quantum field theory in the large {N} limit: a
  review~// {Phys. Rep.} ---
\newblock 2003. ---
\newblock Vol. 385. ---
\newblock Pp.~69--228.

\bibitem{AS95}
\selectlanguageifdefined{english}
{Antonenko~S.~A., Sokolov~A.~I.} Critical exponents for a three-dimensional
  {O}($n$)-symmetrical model with $n>1$~// {Phys.\ Rev. E}. ---
\newblock 1995. ---
\newblock Vol.~51, №~3. ---
\newblock Pp.~1894--1898.

\bibitem{NR82}
\selectlanguageifdefined{english}
{Newman~Kathie~E., Riedel~Eberhard~K.} Cubic $n$-vector model and randomly
  dilute ising model in general dimensions~// {Phys. Rev. B}. ---
\newblock 1982. ---
\newblock Vol.~25, №~1. ---
\newblock Pp.~264--280.

\bibitem{WH73}
\selectlanguageifdefined{english}
{Wegner~Franz~J., Houghton~Anthony}. Renormalization group equation for
  critical phenomena~// {Phys. Rev. A}. ---
\newblock 1973. ---
\newblock Vol.~8, №~1. ---
\newblock Pp.~401--412.

\bibitem{Pinn94}
\selectlanguageifdefined{english}
{Pinn~K., Pordt~A., Wieczerkowski~C.} Computation of hierarchical
  renormalization-group fixed points and their $\varepsilon$-expansions~// {J.
  Stat. Phys.} ---
\newblock 1994. ---
\newblock Vol.~77, no. 5/6. ---
\newblock Pp.~1572--9613.

\bibitem{BBW04}
\selectlanguageifdefined{english}
{Ballhausen~H., Berges~J., Wetterich~C.} Critical phenomena in continuous
  dimension~// {Phys. Lett. B}. ---
\newblock 2004. ---
\newblock Vol. 582, №~2. ---
\newblock Pp.~144--150.

\bibitem{BB01}
\selectlanguageifdefined{english}
{Bagnuls~C., Bervillier~C.} Exact renormalization group equations: an
  introductory review~// {Physics Reports}. ---
\newblock 2001. ---
\newblock Vol. 348, no. 1-2. ---
\newblock Pp.~91--157.

\bibitem{BTW02}
\selectlanguageifdefined{english}
{Berges~J., Tetradis~N., Wetterich~C.} Non-perturbative renormalization flow in
  quantum field theory and statistical physics~// {Physics Reports}. ---
\newblock 2002. ---
\newblock Vol. 363, no. 4-6. ---
\newblock Pp.~223--386.

\bibitem{LZ87}
\selectlanguageifdefined{english}
{Le~Guillou~J.~C., Zinn-Justin~J.} Accurate critical exponents for {I}sing like
  systems in non-integer dimensions~// {J. de Physique}. ---
\newblock 1987. ---
\newblock Vol.~48, №~1. ---
\newblock Pp.~19--25.

\bibitem{DO08}
\selectlanguageifdefined{english}
{O'Dwyer~J., Osborn~H.} Epsilon expansion for multicritical fixed points and
  exact renormalisation group equations~// {Ann. Phys.} ---
\newblock 2008. ---
\newblock Vol. 323, №~8. ---
\newblock Pp.~1859--1898.

\bibitem{Pol84}
\selectlanguageifdefined{english}
{Polchinski~Joseph}. Renormalization and effective lagrangians~// {Nucl. Phys.
  B}. ---
\newblock 1984. ---
\newblock Vol. 231, №~2. ---
\newblock Pp.~269--295.

\bibitem{HH86}
\selectlanguageifdefined{english}
{Hasenfratz~Anna, Hasenfratz~Peter}. Renormalization group study of scalar
  field theories~// {Nucl. Phys. B}. ---
\newblock 1986. ---
\newblock Vol. 270. ---
\newblock Pp.~687--701.

\bibitem{Hol06c}
\selectlanguageifdefined{english}
{Delamotte~B., Holovatch~Y., Ivaneyko~D., Mouhanna~D., Tissier~M.} Spurious
  fixed points in frustrated magnets. ---
\newblock 2006. ---
\newblock arXiv:cond-mat/0609285v1.

\bibitem{Hol06r}
\selectlanguageifdefined{english}
{A~Pelissetto~E~Vicari}. Comment on "{S}purious fixed points in frustrated
  magnets," cond-mat/0609285. ---
\newblock 2006. ---
\newblock arXiv:cond-mat/0610113v1.

\bibitem{Hol06a}
\selectlanguageifdefined{english}
{Delamotte~B., Holovatch~Y., Ivaneyko~D., Mouhanna~D., Tissier~M.} Reply to:
  ``{C}omment on `spurious fixed points in frustrated magnets,'
  cond-mat/0609285". ---
\newblock 2006. ---
\newblock arXiv:cond-mat/0610613v1.

\bibitem{Hol08}
\selectlanguageifdefined{english}
{Delamotte~B, Holovatch~Yu, Ivaneyko~D, Mouhanna~D, Tissier~M}. Fixed points in
  frustrated magnets revisited~// {J. Stat. Mech.} ---
\newblock 2008. ---
\newblock Vol. 2008, №~03. ---
\newblock P.~P03014 (17pp).

\bibitem{BFT93}
\selectlanguageifdefined{english}
{Broadhurst~D.~J., Fleischer~J., Tarasov~O.~V.} Two-loop two-point functions
  with masses: asymptotic expansions and {T}aylor series, in any dimension~//
  {Z. Phys. C}. ---
\newblock 1993. ---
\newblock Vol.~60, №~2. ---
\newblock Pp.~287--301.

\bibitem{Tarasov96}
\selectlanguageifdefined{english}
{Tarasov~O.~V.} Connection between {F}eynman integrals having different values
  of the space-time dimension~// {Phys. Rev. D}. ---
\newblock 1996. ---
\newblock Vol.~54, №~10. ---
\newblock Pp.~6479--6490.

\bibitem{DavDel98}
\selectlanguageifdefined{english}
{Davydychev~A.~I., Delbourgo~R.} A geometrical angle on {F}eynman integrals~//
  {J. Math. Phys.} ---
\newblock 1998. ---
\newblock Vol.~39, №~9. ---
\newblock Pp.~4299--4334.

\bibitem{DavKal01}
\selectlanguageifdefined{english}
{Davydychev~A.~I., Kalmykov~M.~{Yu}.} New results for the
  $\varepsilon$-expansion of certain one-, two- and three-loop {F}eynman
  diagrams~// {Nucl. Phys. B}. ---
\newblock 2001. ---
\newblock Vol. 605, no. 1-3. ---
\newblock Pp.~266--318.

\bibitem{FJT03}
\selectlanguageifdefined{english}
{Fleischer~J., Jegerlehner~F., Tarasov~O.~V.} A new hypergeometric
  representation of one-loop scalar integrals in $d$ dimensions~// {Nucl. Phys.
  B}. ---
\newblock 2003. ---
\newblock Vol. 672, no. 1-2. ---
\newblock Pp.~303--328.

\bibitem{LapRem05}
\selectlanguageifdefined{english}
{Laporta~S., Remiddi~E.} Analytic treatment of the two loop equal mass sunrise
  graph~// {Nucl. Phys. B}. ---
\newblock 2005. ---
\newblock Vol. 704, no. 1-2. ---
\newblock Pp.~349--386.

\bibitem{Tarasov06}
\selectlanguageifdefined{english}
{Tarasov~O.~V.} Hypergeometric representation of the two-loop equal mass
  sunrise diagram~// {Phys. Lett. B}. ---
\newblock 2006. ---
\newblock Vol. 638, no. 2-3. ---
\newblock Pp.~195--201.

\bibitem{ArgMast07}
\selectlanguageifdefined{english}
{Argeri~M., Mastrolia~P.} Feynman diagrams and differential equations~// {Int.
  J. Mod. Phys. A}. ---
\newblock 2007. ---
\newblock Vol.~22, №~24. ---
\newblock Pp.~4375 -- 4436.

\bibitem{GKP07}
\selectlanguageifdefined{english}
{Groote~S., K{\"o}rner~J.~G., Pivovarov~A.~A.} On the evaluation of a certain
  class of {F}eynman diagrams in $x$-space: {S}unrise-type topologies at any
  loop order~// {Annals of Physics}. ---
\newblock 2007. ---
\newblock Vol. 322, №~10. ---
\newblock Pp.~2374--2445.

\bibitem{Sh07}
\selectlanguageifdefined{english}
{Shpot~M.~A.} A massive {F}eynman integral and some reduction relations for
  {A}ppell functions~// {J. Math. Phys.} ---
\newblock 2007. ---
\newblock Vol.~48, №~12. ---
\newblock Pp.~123512--1---13.

\bibitem{Tarasov08}
\selectlanguageifdefined{english}
{Tarasov~O.~V.} New relationships between {F}eynman integrals~// {Phys. Lett.
  B}. ---
\newblock 2008. ---
\newblock Vol. 670. ---
\newblock Pp.~67--72.

\bibitem{KK09}
\selectlanguageifdefined{english}
{Kalmykov~Mikhail~Yu., Kniehl~Bernd~A.} Towards all-order laurent expansion of
  generalised hypergeometric functions about rational values of parameters~//
  {Nucl. Phys. B}. ---
\newblock 2009. ---
\newblock Vol. 809, №~3. ---
\newblock Pp.~365 -- 405.

\bibitem{Callan70}
\selectlanguageifdefined{english}
{Callan~Curtis~G.} Broken scale invariance in scalar field theory~// {Phys.
  Rev. D}. ---
\newblock 1970. ---
\newblock Vol.~2, №~8. ---
\newblock Pp.~1541--1547.

\bibitem{Sym70}
\selectlanguageifdefined{english}
{Symanzik~K.} Small distance behaviour in field theory and power counting~//
  {Comm. Math. Phys.} ---
\newblock 1970. ---
\newblock Vol.~18, №~3. ---
\newblock Pp.~227--246.

\bibitem{Sym71}
\selectlanguageifdefined{english}
{Symanzik~K.} Small-distance-behaviour analysis and {W}ilson expansions~//
  {Comm. Math. Phys.} ---
\newblock 1971. ---
\newblock Vol.~23, №~1. ---
\newblock Pp.~48--86.

\bibitem{BLZ73}
\selectlanguageifdefined{english}
{Brezin~E., Guillou~Le, C.~J., Zinn-Justin~J.} Wilson's theory of critical
  phenomena and {C}allan-{S}ymanzik equations in $4-\epsilon$ dimensions~//
  {Phys. Rev. D}. ---
\newblock 1973. ---
\newblock Vol.~8, №~2. ---
\newblock Pp.~434--440.

\bibitem{Jug83}
\selectlanguageifdefined{english}
{Jug~G.} Critical behavior of disordered spin systems in two and three
  dimensions~// {Phys. Rev. B}. ---
\newblock 1983. ---
\newblock Vol.~27, №~1. ---
\newblock Pp.~609--612.

\bibitem{BNGM76}
\selectlanguageifdefined{english}
{Baker~George~A., Nickel~Bernie~G., Green~Melville~S., Meiron~Daniel~I.}
  Ising-model critical indices in three dimensions from the {C}allan-{S}ymanzik
  equation~// {Phys. Rev. Lett.} ---
\newblock 1976. ---
\newblock Vol.~36, №~23. ---
\newblock Pp.~1351--1354.

\bibitem{NMB77}
\selectlanguageifdefined{english}
{Nickel~B.~G., Meiron~D.~I., G.~A.~Baker~Jr.} Compilation of 2-pt and 4-pt
  graphs for continuum spin models. ---
\newblock 1977. ---
\newblock Guelph University Report. {\small\url{
  http://www.citebase.org/abstract?id=oai:arXiv.org:cond-mat/0610113}}.

\bibitem{AS}
\selectlanguageifdefined{english}
{Abramowitz~Milton, Stegun~Irene~A.} Handbook of Mathematical Functions. ---
\newblock New York: National Bureau of Standards, 1972.

\bibitem{Ami84}
\selectlanguageifdefined{english}
{Amit~Daniel~J.} Field theory, the renormalization group, and critical
  phenomena. ---
\newblock Singapore: World Scientific, 1984, 2nd edition.

\bibitem{PBM3}
\selectlanguageifdefined{english}
{Prudnikov~A., Brychkov~Yu.~A., Marichev~O.~I.} Integrals and Series. More
  special functions. ---
\newblock New York: Gordon and Breach, 1990. ---
\newblock Vol.~3.

\bibitem{Lewin}
\selectlanguageifdefined{english}
{Lewin~L.} Polylogarithms and associated functions. ---
\newblock New York: Elsevier, 1981.

\bibitem{BD91}
\selectlanguageifdefined{english}
{Boos~E.~E., Davydychev~A.~I.} A method of calculating massive {F}eynman
  integrals~// {Teor. Mat. Fiz.} ---
\newblock 1991. ---
\newblock Vol.~89. ---
\newblock Pp.~56--72. ---
\newblock [Sov. Phys. Theor. Math. Phys. {\bf 89}, 1052--1064 (1992)].

\bibitem{PV00np}
\selectlanguageifdefined{english}
{Pelissetto~Andrea, Vicari~Ettore}. The effective potential of {N}-vector
  models: a field-theoretic study to ${O}(\varepsilon^3)$~// {Nucl. Phys. B}.
  ---
\newblock 2000. ---
\newblock Vol. 575, №~3. ---
\newblock Pp.~579--598.

\bibitem{DavTausk96}
\selectlanguageifdefined{english}
{Davydychev~A.~I., Tausk~J.~B.} Connection between certain massive and massless
  diagrams~// {Phys. Rev. D}. ---
\newblock 1996. ---
\newblock Vol.~53, №~12. ---
\newblock Pp.~7381--7384.

\bibitem{math}
\selectlanguageifdefined{english}
Mathematica 4.1, a product of Wolfram Research.

\bibitem{PBM1}
\selectlanguageifdefined{english}
{Prudnikov~A., Brychkov~Yu.~A., Marichev~O.~I.} Integrals and Series.
  Elementary functions. ---
\newblock New York: Gordon and Breach, 1986. ---
\newblock Vol.~1.

\bibitem{HK94}
\selectlanguageifdefined{english}
{Holovatch~Yu., Krokhmal's'kii~T.} Compilation of two-point and four-point
  graphs in field theory in noninteger dimensions~// {J. Math. Phys.} ---
\newblock 1994. ---
\newblock Vol.~35, №~8. ---
\newblock Pp.~3866--3880.

\end{thebibliography}

\ukrainianpart

\title{Двопетлеві РГ-функції масивної теорії поля $\bm{\phi^4}$ у
довільних вимірностях простору}
\author{М .А. Шпот}
\address{Інститут Фізики Конденсованих Систем НАН України, 79011 Львів, Україна}
\makeukrtitle
\begin{abstract}
\tolerance=3000%
Двопетлеві інтеграли Фейнмана масивної теорії поля типу $\phi^4_d$
розраховані в явному вигляді для довільних вимірностей простору $d$.
Відповідні ренормгрупові функції записані у компактному вигляді
у термінах Гаусових гіпергеометричних функцій.
Продемонстровано ряд цікавих та корисних
співвідношень для цих інтегралів а також для деяких спеціальних
математичних функцій та констант.
\keywords Інтеграли Фейнмана, спеціальні функції, ренормалізаційна група, теорія поля
\pacs 11.10.Kk, 12.38.Bx, 02.30.Gp, 05.70.Jk
\end{abstract}

\end{document}